\documentclass[preprint, amsmath, amssymb, aps]{revtex4-1}

\usepackage{graphicx}
\usepackage{dcolumn}
\usepackage{bm}
\usepackage{tensor}

\begin{document}

\title{Medium Generated Gap in Gravity and a 3d Gauge Theory}

\author{Gregory Gabadadze}
\author{Daniel Older}
\affiliation{
Center for Cosmology and Particle Physics, Department of Physics,\\
New York University, New York, NY 10003, USA
}

\date{\today}

\begin{abstract}
It is well known that a physical medium that sets a Lorentz frame generates a Lorentz-breaking 
gap for a graviton.  We examine such  generated "mass" terms  in the presence of a fluid  
medium whose ground state spontaneously breaks spatial translation invariance 
in $d = D+1$ spacetime dimensions, and for a solid in $D = 2$ spatial dimensions.  
By requiring energy positivity and subluminal propagation, certain constraints are placed on the equation of state of the 
medium.  In the case of $D = 2$ spatial dimensions, classical gravity can be recast as a Chern-Simons gauge theory and motivated by this we recast the massive  theory of gravity in AdS$_3$ as a massive Chern-Simons gauge theory with an unusual mass term.  We find that in the flat space limit the Chern-Simons theory has a  novel 
gauge invariance that mixes the kinetic and mass terms, and enables the massive theory with 
a non-compact internal group to be free of ghosts and  tachyons.

\end{abstract}

\maketitle

\section{\label{sec:level1} Introduction and Summary}
It is well known that massless spin 1 bosons can dynamically acquire a mass at low energies.
This occurs when the Nambu-Goldstone (NG) bosons non-linearly realizing  a global symmetry
become longitudinal modes of massive gauge bosons, after the gauge fields are introduced.
Two long known examples are of a photon acquiring  a mass in a neutral plasma \cite{Anderson1963}, 
where there is a massless remnant,  and of the Brout-Englert-Higgs mechanism 
\cite{EnglertBrout1964, Higgs1964, Kibble1964}, where the NG's are absorbed into the gauge 
fields while the remaining non-NG massive boson can still  be light at weak coupling.

A natural question to ask is whether  similar mechanisms can work in the case of a massless spin-2 field.
Lorentz-invariant models  will not be the subject of our work, instead we ask whether gravity can dynamically acquire a 
gap via coupling to matter that sets a preferred rest frame?  A  well known example is of  
the Jeans instability due to a longitudinal graviton acquiring a tachyonic mass
within a (nearly) pressure-less matter density distribution. The tachyon instability in this case describes  
a collapse of the (nearly) pressure-less matter by attractive gravity. 
Here, we study a more general case in which a fluid or solid medium with unspecified equation of state described by D scalar fields (the comoving with matter coordinates) is minimally coupled to gravity and whose rest frame  
sets a preferred frame. In describing such a continuous medium, we follow closely the notations and conventions of \cite{Soper2008, Dubovsky2005, Endlich2011}. The NG bosons for this medium are the longitudinal and transverse phonons that realize non linearly  spatial translation invariance. In the unitary gauge, the graviton swallows all the phonons and acquires general Lorentz violating mass terms.

For a perfect fluid in any dimension, the low energy action in unitary gauge gives rise to propagating tensor (in $D \ge 3$), vector (in $D \ge 2$), and scalar modes. For an elastic solid in D = 2 spatial dimensions, one vector and one scalar degree of freedom propagate. In both cases, we find that imposing conditions of positive energy density and sub-luminal propagation places certain constraints on the equation of state of the medium. Additionally, we confirm at the non-linear level that there are no ghosts in the theory which is naturally expected since the theory is merely ordinary continuous matter coupled to gravity. 

In 3 spacetime dimensions, gravity has a dual description as a Chern-Simons gauge theory with non-compact gauge group \cite{Achucarro1986, WITTEN1986}. When a fluid or solid medium couples to gravity, the low energy dynamics of this theory can also be viewed, in the weakly coupled perturbative regime, as the dynamics of a Chern-Simons gauge field with an acquired mass term. We investigate this theory in the special case of gravity in an AdS$_3$ background in which the acquired mass term is none other than the usual Fierz-Pauli mass term. As a gauge theory however, this mass term appears quite unusual and to our knowledge has not been introduced in the literature. We find that it has an exact gauge redundancy the  form of which appears new and is a consequence of the particular combination of the Chern-Simons kinetic term and unusual mass term appearing together in the Lagrangian. We investigate the dynamics of such a theory and find that it indeed propagates 2 degrees of freedom as we expect from the dual description in terms of the metric.

Such a system that gives rise to an infrared modification of gravity is of interest for its potential applications in cosmology and condensed matter physics. In cosmology, the large scale structure is determined precisely by the interaction between gravity and a cosmic fluid medium. Gravity would generate massive Lorentz violating excitations from perturbations of the fluid in a similar manner to the mechanism proposed in this paper.  At very large scales such massive gravity description 
should be preferred over the one that treats separately  massless gravitons from  fluid fluctuations, more appropriate 
at shorter scales. On the condensed matter side, it is well-known that the collective behavior in certain many-body  systems are well described by the properties of both Abelian and non-Abelian gauge fields with Chern-Simons terms and in some cases, these Chern-Simons fields can have effective mass terms \cite{Dunne1999}.  The massive Chern-Simons gauge theory we study here has not been introduced as of yet in the literature but could plausibly arise in some condensed matter system yet to be studied. In this paper, we defer such applications however, and focus mainly on the results for the dynamics and stability of such systems.

The paper is organized as follows. In Section 2, we overview  the most general Lorentz violating massive gravity theory at the linearized level, looking at the dynamics and matter coupling. In Section 3, we look at a material medium minimally coupled to gravity and understand the low energy linearized dynamics of the theory in the presence and absence of gravity and what sector of Lorentz violating massive gravity the linearized theory falls under. Finally, in Section 4 we look at the linearized dynamics of Chern-Simons Fierz-Pauli gravity in $D = 2$. 

\section{\label{sec:level1} Lorentz Violating Massive Gravity}

We begin by reviewing the propagating degrees of freedom for different sectors of the mass parameter space in a generic Lorentz-violating massive gravity theory. Foundational work has already  been done on this previously \cite{Rubakov2004, Dubovsky2004} for broad sectors of mass-parameter space in D = 3, but we would like to extend the analysis to arbitrary spatial dimensions and analyze both the free field dynamics and linear response to a conserved energy-momentum tensor. We start with linearized Einstein gravity in $(D+1)$-Minkowski spacetime $\mathcal{L}_0$ 
plus a generic Lorentz violating mass term $\mathcal{L}_m$:
\begin{align} \label{LVFP}
\mathcal{L}_0 &= -\frac{1}{2} \partial_{\lambda} h_{\mu \nu} \partial^{\lambda} h^{\mu \nu} + \partial_{\mu} h_{\lambda \nu} \partial^{\nu} h^{\lambda \mu} - \partial_{\mu} h^{\mu \nu} \partial_{\nu} h + \frac{1}{2} \partial_{\lambda} h \partial^{\lambda} h \,,\nonumber \\
\mathcal{L}_m &= m_0^2 h_{00}^2 +m_1^2 h_{0i}^2 -\frac{1}{2} m_2^2 h_{ij}^2 + \frac{1}{2} m_3^2 \hat{h}^2 - m_4^2 h_{00} \hat{h}\,,
\end{align}
where $\hat{h} \equiv \delta^{ij} h_{ij}$ and $i = 1, \dots, D$.   The above Lagrangian requires
clarifications.  As written,  it  assumes the background spacetime to be that of Minkowski, while the fluctuations are presumed to be Lorentz-violating in general.  In conventional cases, however, 
when a conventional matter source sets a preferred frame, it  would also back-react on  spacetime to   take it away  
from a flat Minkowski, and hence, the above Lagrangian might appear to be inappropriate. 

Nevertheless, at energies and momenta higher than a  typical  curvature scale set by the  matter source, the background 
curvature won't be important, and hence  the above Lagrangian can be used to count and characterize  
approximately the degrees of freedom.  This will be the approach  in  the present section. There is one caveat though related to the fact  that  this approach ignores the  constant and linear
terms  that will also exist  when  Minkowski spacetime is  not an exact  
solution. Do these terms matter at high energies/momenta?  They do  only in special cases,  when 
ignoring the linear  terms confuses what actually is a
Lagrange multiplier field with an algebraically determined field;   as a result,  one is led to a naive and incorrect 
counting of the degrees of freedom. This is discussed in  the next section where the full dynamics of the source 
is accounted for and the constant and linear terms are  included.    Till the next section we overview and extend 
the analysis of the degrees of freedom in (\ref {LVFP}). The reader who's 
familiar with this part can skip directly to the next section. 
  
Generally speaking, we can split metric perturbations up into tensors, vectors, and scalars as follows:
\begin{align}
h_{00} &= \psi \,,\nonumber \\
h_{0i} &= u_i + \partial_i v \,,\nonumber \\
h_{ij} &= \chi_{ij} + \partial_i s_j + \partial_j s_i + \partial_i \partial_j \sigma + \delta_{ij} \tau.
\end{align}
where $u_i$ and $s_i$ are transverse vectors and $\chi_{ij}$ is a transverse, traceless, symmetric tensor. The Lagrangian naturally splits up into scalar, vector and tensor sectors. Additionally, we can add a coupling to a conserved energy momentum tensor $h_{\mu \nu} T^{\mu \nu}$ and determine how the matter sources the fields in the tensor, vector, and scalar sectors. We split up our energy-momentum tensor as follows:
\begin{align}
T_{00} &= \rho\,, \nonumber \\ 
T_{0 i} &= p_i + \partial_i f\,, \nonumber \\
T_{i j} &= \Sigma_{i j} + \partial_i \phi_j + \partial_j \phi_i + \partial_i \partial_j b + \delta_{i j} T\,,
\end{align}
where, mirroring the decomposition for $h_{\mu \nu}$, we have that $p_i$ and $\phi_i$ are transverse, and $\Sigma_{i j}$ is transverse, traceless and symmetric. Additionally, since our energy-momentum tensor is conserved, we have three conditions that must be satisfied:
\begin{align}
\partial_0 \rho &= \nabla^2 f \,,\nonumber \\
\partial_0 p_i &= \nabla^2 \phi_i\,, \nonumber \\
\partial_0 \partial_i f &= \partial_i \nabla^2 b + \partial_i T\,,
\end{align}
making our source term in the Lagrangian to look as follows:
\begin{align}
\mathcal{L}_{source} = h_{\mu \nu} T^{\mu \nu} = \psi \rho - 2 u_i p_i + 2 v \partial_0 \rho -2 s_i \partial_0 p_i + \sigma \partial_0^2 \rho + \tau (\nabla^2 b + D T) + \chi_{i j} \Sigma_{i j}\,. 
\end{align}
We shall analyze the dynamics and stability of each of these sectors without the source term and the solutions for the fields with generic source term in what follows.

\subsection{\label{sec:level2} Tensor Sector}

The tensor sector is the simplest in Lorentz violating massive gravity. When $D = 1, 2$,  no dynamical transverse, traceless, symmetric tensors exist for such small dimensions. 
For $D \ge 3$, the free theory gives us: 
\begin{align}
\mathcal{L} = \frac{1}{2} \chi_{i j} \Big[ -\partial_0^2 + \nabla^2 - m_2^2 \Big] \chi_{i j}.
\end{align}
Clearly, from this Lagrangian, we see that the transverse, traceless symmetric tensor propagates $\frac{(D+1)(D-2)}{2}$ massive degrees of freedom with mass $m_2$ and that the Hamiltonian will be positive definite. 

Adding the source term, we get that the generic solution to the graviton tensor modes in $D \ge 3$ are:
\begin{align}
\chi_{i j} = \frac{-1}{\Box - m_2^2} \Sigma_{i j}\,.
\end{align}

\subsection{\label{sec:level2} Vector Sector}

For the vector sector, we get the following Lagrangian for the free theory:
\begin{align}
\mathcal{L}_V = s_j (\partial_0^2 \nabla^2 + m_2^2 \nabla^2) s_j -u_j (\nabla^2 - m_1^2) u_j + 2 u_j \nabla^2 \partial_0 s_j\,.
\end{align}
Note that $u_i$ can be integrated out, and then the equations of motion for 
$s_i$ give us the following dispersion relation:
\begin{align}
\omega^2 = \frac{m_2^2}{m_1^2} k^2 + m_2^2\,.
\end{align}
Thus, we have $D-1$ propagating degrees of freedom which have a normal massive dispersion relation in the case when the masses are equal. In the case where either $m_2 = 0$ or $m_1 = 0$, the vector modes carry no propagating degrees of freedom. Finally, in the massless limit while keeping the ratio $\frac{m_2}{m_1}$ fixed, we get a massless free vector field which also agrees with the massless limit of Fierz-Pauli theory. The only condition that this dispersion relation imposes upon us is that 
$m_1 \ge m_2$ so that there is no superluminal propagation.

Additionally, the Hamiltonian for the vector sector is:
\begin{align}
\mathcal{H} = \frac{1}{4} \pi_i \Big( \frac{\nabla^2 - m_1^2}{m_1^2 \nabla^2}\Big) \pi_i -s_i (m_2^2 \nabla^2) s_i\,,
\end{align}
where $\pi_i$ is the conjugate momentum to $s_i$. This is clearly positive, or zero, 
which shows that our propagating vector mode is stable. 

If we add a source term, we get that the generic solution for the vector modes are:
\begin{align}
u_i &= \frac{-\partial_0^2 \nabla^2 p_i}{(\nabla^2 - m_1^2)(-\partial_0^2 \nabla^2 + \frac{m_2^2}{m_1^2} \nabla^2 \nabla^2 - m_2^2 \nabla^2)}\,,\nonumber \\
s_i &= \frac{-\partial_0 p_i}{-\partial_0^2 \nabla^2 + \frac{m_2^2}{m_1^2} \nabla^2 \nabla^2 - m_2^2 \nabla^2}\,.
\end{align}

\subsection{\label{sec:level2} Scalar Sector}

For the scalar sector, the Lagrangian is:
\begin{align}
\mathcal{L}_S = &\frac{1}{2} \tau \Big[ (D^2 - D) \partial_0^2 -(D^2 - 3 D + 2) \nabla^2 + D^2 m_3^2 - D m_2^2 \Big] \tau + \sigma \Big( \frac{1}{2} (m_3^2 - m_2^2)\nabla^2 \nabla^2 \Big) \sigma \nonumber \\ 
+ &m_0^2 \psi^2 - v \Big(m_1^2 \nabla^2 \Big) v + \tau \Big( (D-1) \partial_0^2 \nabla^2 + (D m_3^2 - m_2^2)\nabla^2 \Big) \sigma \nonumber \\ 
+ &\tau \Big( (D-1) \nabla^2 - D m_4^2\Big) \psi - \tau \Big( 2 (D-1) \nabla^2 \partial_0 \Big) v -\psi \Big( m_4^2 \nabla^2 \Big) \sigma. 
\end{align}
Not all independent choices for the five masses will give a sensible physical theory so we must analyze the dispersion relations and Hamiltonian for different sectors of the mass parameter space.

{\subsubsection*{Case 1: \label{sec:level2} $\mathbf{m_0 \ne 0, m_1 \ne 0}$}}

In this case, we can integrate out $v$ and then $\psi$ giving us a Lagrangian in terms of $\tau$ and $\sigma$. The kinetic part of the Lagrangian is then:
\begin{align}
\mathcal{L}_{kin} &= \tau \Big( \partial_0^2 -\frac{(D-1)^2}{m_1^2} \partial_0^2 \nabla^2 \Big) \tau + \tau \Big( (D-1) \partial_0^2 \nabla^2 \Big) \sigma\,.
\end{align}
This theory has a ghost in the spectrum and is thus, unstable. We therefore disregard this sector and forego the need to determine the solutions for a source.

\subsubsection*{Case 2: \label{sec:level2}  $\mathbf{m_0 \ne 0, m_1 = 0}$}

For this case in the free theory, the equations of motion for $v$ force $\tau$ to be time independent. This dissolves any time dependence from the other fields and consequently, there are no freely propagating degrees of freedom. 

Upon adding a source term, we get the general solution for the scalar modes as:
\begin{align}
\tau &= -\frac{1}{ (D-1) \nabla^2} \rho \,, \nonumber \\
\sigma &= \frac{D m_3^2 - m_2^2 - D \frac{m_4^4}{m_0^2}}{(D-1)(m_3^2 - m_2^2 - \frac{m_4^4}{2 m_0^2})\nabla^2 \nabla^2} \rho \,, \nonumber \\
\psi &= \Big[\frac{D m_4^2}{2(D-1) m_0^2 \nabla^2} - \Big(\frac{m_4^2}{2 m_0^2}\Big) \frac{D m_3^2 - m_2^2 - D \frac{m_4^4}{m_0^2}}{(D-1)(m_3^2 - m_2^2 - \frac{m_4^4}{2 m_0^2}) \nabla^2} \Big] \rho \,, \nonumber \\
v &= \Big( -\frac{D}{2 (D-1)} \frac{\partial_0}{\nabla^4} +\frac{(D-2)}{2 (D-1)}\frac{1}{\nabla^2 \partial_0} - \frac{D^2 m_3^2 - D m_2^2}{2 (D-1)^2 \nabla^4 \partial_0} \Big) \rho + \frac{\nabla^2 b + D T}{2 (D-1) \nabla^2 \partial_0}\,.
\end{align}
These solutions describe an instantaneous response of a field to a source, and aren't physical.

\subsubsection*{Case 3: \label{sec:level2} $\mathbf{m_0 = 0, m_4 = 0}$}

For this case, $\psi$ acts as a Lagrange multiplier enforcing $\nabla^2 \tau = 0$ which means for fields that die away at spatial infinity that $\tau = 0$, so our free fields are all time independent and don't propagate. 

When we couple the  theory to a conserved energy-momentum tensor, we find the scalar modes to be:
\begin{align}
\tau &= -\frac{1}{ (D-1) \nabla^2} \rho \, \nonumber \\
\sigma &= \frac{Dm_3^2 - m_2^2}{(D-1) (m_3^2 - m_2^2) \nabla^2 \nabla^2} \rho \,, \nonumber \\
\psi &= \Big[ \frac{D}{D-1} \frac{\partial_0^2}{\nabla^4} -\frac{D-2}{D-1} \frac{1}{\nabla^2} + \frac{D^2 m_3^2 - D m_2^2}{(D-1)^2} \frac{1}{\nabla^4} \Big] \rho + \frac{\nabla^2 b + D T}{(D-1) \nabla^2} \,, \nonumber \\
v &= \frac{4 \partial_0}{m_1^2 \nabla^2} \rho \,.
\end{align}

\subsubsection*{Case 4: \label{sec:level2} $\mathbf{m_0 = 0, m_4 \ne 0, m_1 = 0}$}

For this case, both $\psi$ and $v$ are Lagrange multipliers and the equations of motion give us 
a trivial theory where the fields vanish uniformly so long as they die off at spatial infinity. 

When we couple our theory to matter however, we get:
\begin{align}
\tau &= -\frac{1}{ (D-1) \nabla^2} \rho\,, \nonumber \\
\sigma &= \frac{1}{m_4^2 \nabla^2} \rho\,, \nonumber \\
\psi &=  \Big[ \frac{m_3^2 - m_2^2}{m_4^4} - \frac{D m_3^2 - m_2^2}{(D-1) m_4^2} \frac{1}{\nabla^2}\Big] \rho \,, \nonumber \\
v &= -\Big[\frac{\partial_0}{2 m_4^2 \nabla^2} - \frac{D}{2 (D-1)} \frac{\partial_0}{\nabla^4} + \frac{m_3^2 - m_2^2}{2 m_4^4} \frac{1}{\partial_0} + \Big( \frac{D-2}{2 (D - 1)} - \frac{D (m_3^2 - m_2^2)}{2 (D-1) m_4^2} \Big) \frac{1}{\nabla^2 \partial_0} \Big] \rho  \nonumber  \\ &+ \frac{\nabla^2 b + DT}{2 (D-1) \nabla^2 \partial_0} \,. \nonumber
\end{align}

\subsubsection*{Case 5: \label{sec:level2} $\mathbf{m_0 = 0, m_4 \ne 0, m_1 \ne 0}$}

For $m_0 = 0$ and $m_1, m_4 \ne 0$, we first consider the free field case. We can integrate out all of the fields except $\tau$ and  get the following Lagrangian:
\begin{align}
\mathcal{L} = &\tau \Big( -\frac{D (D-1)}{2} \partial_0^2 + [(D - 1)^2 \Big(\frac{m_2}{m_4}\Big)^2 - \frac{D^2 - 3D + 2}{2}] \nabla^2 + \Big(\frac{(D - 1)^2 (m_1^2 - m_4^2)}{m_1^2 m_4^2}\Big) \partial_0^2 \nabla^2 \nonumber \\
+ &\Big( \frac{(D-1)^2 (m_3^2 - m_2^2)}{2 m_4^2} \Big) \nabla^2 \nabla^2 -\frac{D (D-1)}{2} m_2^2  \Big) \tau\,.
\end{align}
The requirement  that the  Hamiltonian be bounded from below puts bounds on what our masses can be. Specifically, for ensure kinetic and potential energy cannot become arbitrarily negative, we need:
\begin{align}
m_1^2 &\ge m_4^2 \label{poskin} \,,\\
m_2^2 &\ge m_3^2 \label{pospot}\,.
\end{align}
The dispersion relation we get for $\tau$ is:
\begin{align} \label{dispersion}
\omega^2 = \frac{2 (D-1)^2 (\frac{m_2^2 - m_3^2}{2 m_4^4})k^4 + 2 [(D - 1)^2(\frac{m_2}{m_4})^2 - \frac{D^2 - 3D + 2}{2}] k^2 + D (D-1) m_2^2}{D (D-1)  + 2 (D - 1)^2 (\frac{m_1^2 - m_4^2}{m_1^2 m_4^2}) k^2}\,.
\end{align}
We don't allow the  scalar mode to go faster than the speed of light so we require, conservatively,  that the group velocity is less than 1. Since the second derivative of $\omega (k)$ is positive, we need only look at the group velocity 
at large k in which case we get:
\begin{align}
\omega(k) \sim \frac{m_1^2 (m_2^2 - m_3^2)}{2 m_4^2 (m_1^2 - m_4^2)} k^2\,,
\end{align}
which gives us the additional condition that:
\begin{align}
m_1^2 (m_2^2 - m_3^2) \le 2 m_4^2 (m_1^2 - m_4^2)\,. \label{sublum}
\end{align}
If this condition is satisfied, then we will have no superluminal propagation of the scalar mode.

Looking at the special case where the mass term is the Lorentz invariant Fierz-Pauli mass term (i.e. $m_0 = 0, m_i = m$ for $i = 1,2,3,4$), the dispersion relation \eqref{dispersion} reduces to:
\begin{align}
\omega^2 = k^2 + m^2\,,
\end{align}
in any dimension D which is in agreement with what we expect that Fierz-Pauli carries one massive scalar degree of freedom.

When we couple our theory to matter, we get:
\begin{align}
\tau &= \frac{(-2\frac{m_1^2- m_4^2}{(D-1) m_4^2} \partial_0^2 + \frac{D m_1^2}{(D-1)^2} \frac{\partial_0^2}{\nabla^2} - \frac{m_1^2 (m_3^2 - m_2^2)}{(D-1) m_4^4} \nabla^2 - \frac{m_1^2 m_2^2}{(D-1) m_4^2}) \rho - \frac{m_1^2}{(D-1)^2} (\nabla^2 b + D T)}{-\frac{D}{D-1} m_1^2 \partial_0^2 + (m_3^2 - m_2^2) (\frac{m_1^2}{m_4^2}) \nabla^2 \nabla^2 + 2 (1 +(\frac{m_1^2}{m_4^2}) ) \nabla^2 \partial_0^2 + (2 (\frac{m_2^2}{m_4^2}) - \frac{D-2}{D-1}) \nabla^2 - \frac{D}{D-1} m_2^2} \,,\nonumber \\
\sigma &= \frac{1}{m_4^2 \nabla^2} \rho + (\frac{D-1}{m_4^2} - \frac{D}{\nabla^2}) \tau \,, \nonumber \\
\psi &=  \frac{(m_3^2 - m_2^2) \nabla^2}{m_4^2} \sigma + \frac{(D-1) \partial_0^2 + D m_3^2 - m_2^2}{m_4^2} \tau + \frac{\partial_0^2}{m_4^2 \nabla^2} \rho \,, \nonumber \\
v &=  \frac{\partial_0}{m_1^2 \nabla^2} \rho - \frac{(D-1) \partial_0}{m_1^2} \tau\,.
\end{align}
Here, for the sake of space we have solved for $\tau$ explicitly and left the solutions for the other fields in terms of $\rho, \tau,$ and $\sigma$.

\section{\label{sec:level1} Mass Generation in Fluids and Solids}

Such Lorentz violating mass terms as we just analyzed can arise when we couple the gravitational field to a continuous matter system whose degrees of freedom are parametrized by co-moving coordinates $\phi^a, a = 1, \dots, D$. We follow a similar notation to that used in \cite{Endlich2011}. Any continuous matter Lagrangian which is homogeneous and couples minimally to the metric can be constructed out of the components of the deformation matrix:
\begin{align}
M^{a b} \equiv g^{\mu \nu} \partial_{\mu} \phi^a \partial_{\nu} \phi^b.
\end{align}
Furthermore, for our medium to be isotropic, the matter Lagrangians can be constructed only out of rotational invariants of the deformation matrix \cite{Soper2008}. In D dimensions, a basis for such invariants are $tr(M^n), n = 1, \dots, D$. Another basis for such invariants, which we will use in this paper, are the principal invariants Det$_n$ for a $D \times D$ matrix defined as follows:
\begin{align}
\text{Det}_n (M_{a b}) \equiv \frac{1}{n! (D-n)!} \epsilon^{a_1 \dots a_D} \epsilon^{b_1 \dots b_D} M_{a_1 b_1} \dots M_{a_n b_n} \delta_{a_{n+1} b_{n+1}} \dots \delta_{a_D b_D}.
\end{align}
These invariants naturally arise in the characteristic polynomial for a $D \times D$ matrix as the coefficients. If the characteristic polynomial for the matrix $M_{a b}$ is $a_0 x^D + a_1 x^{D-1} + \dots + a_{D-1} x + a_D$, then $a_n = (-1)^{D-n} \text{Det}_n (M)$. For a general homogeneous and isotropic material coupled to gravity in (D+1)-dimensional spacetime, the Lagrangian would then be:
\begin{align} \label{Lagrangian}
\mathcal{L} = \sqrt{-g} R + \sqrt{-g} f(\text{Det}_1(M), \dots, \text{Det}_D(M))\,,
\end{align}
where f is some function of the invariants. The specific form of the function f then will determine the equation of state for the medium (note the similarities with and differences from  the Lorentz-invariant theory of massive gravity, \cite {dRG,dRGT}). For example, in D = 2, the two invariants would be:
\begin{align}
\text{Det}_1 (M) &\equiv tr(M) = M^{a b} \delta_{a b} \,,\\
\text{Det}_2 (M) &\equiv \frac{1}{2} (tr(M)^2 - tr(M^2)) = det(M).
\end{align}
We will only be considering the specific cases of perfect fluids in any dimension D and homogeneous and 
isotropic solids in dimension $D = 2$ in this paper. For fluids, the only invariant that the Lagrangian would depend on is the true determinant of the matrix Det$_D$, which we will denote by y. For isotropic solids in 2 dimensions, we will denote the two invariants Det$_1$ and Det$_2$ as x and y, respectively. 
 
The Lagrangian \eqref{Lagrangian} is naturally invariant under general coordinate transformations since every component of the deformation matrix is a scalar field. In addition to this local symmetry, another symmetry of our Lagrangian is translations   and a global $SO(D)$ in the comoving coordinates (in the "$\phi$-space") which follows from homogeneity and isotropy of the medium. This is the largest global symmetry group of the comoving coordinates if our Lagrangian depends generally on invariants of the deformation matrix; the  material system in such a case is generally some type of homogeneous and isotropic elastic solid. On the otherhand, if the Lagrangian only depends on the determinant of the deformation matrix (as is the case with the perfect fluid to be discussed below) then our symmetry group expands considerably from a global SO(D) to all volume preserving diffeomorphisms of the target space of comoving coordinates:
\begin{align}
\phi^a \rightarrow \hat{\phi}^a (\phi) \quad \text{ such that } \quad det(\frac{\partial \hat{\phi}^a}{\partial \phi^b}) = 1.
\end{align}
We take the rest frame of the material medium to be the state in which the comoving coordinates coincide with the real spatial coordinates of Minkowski spacetime. When we expand the metric and fields $\phi^a$ about Minkowski spacetime, we define perturbations $\pi^a, h_{\mu \nu},$  as follows:
\begin{align}
\phi^a &= x^a + \pi^a\,, \\
g_{\mu \nu} &= \eta_{\mu \nu} + h_{\mu \nu}\,.
\end{align}
Under general spatial coordinate transformations $x^a \rightarrow f^a(x) \approx x^a + \xi^a$, our perturbations transform to first order as:
\begin{align}
\delta h_{00} &= 0 \,, \qquad \qquad \qquad \qquad \delta \pi_i = -\xi_i  \,, \nonumber \\
\delta h_{0 i} &= -\partial_0 \xi_i\,,   \qquad \qquad \qquad
\delta h_{i j} = -\partial_i \xi_j - \partial_j \xi_i\,.
\end{align}
In unitary gauge, where $\pi^a \equiv 0$, we have a Lagrangian of the form \eqref{LVFP} up to quadratic order in the metric perturbations with additional linear terms,  and a vacuum constant (the energy density of the medium).

In the absence of gravity, the deformation matrix  and the invariants to quartic order in the $\pi^a$ fields are:
\begin{align}
M_{a b} &= \delta_{a b} + \partial_a \pi_b + \partial_b \pi_a - \partial_0 \pi_a \partial_0 \pi_b + \partial_i \pi_a \partial_i \pi_b \,, \\
x &= D + 2 \partial_i \pi_i - (\partial_0 \pi_i)^2 + (\partial_i \pi_j)^2 \,, \\
y &= 1 + 2(\partial_i \pi_i) - (\partial_0 \pi_i)^2 + (\partial_i \pi_i)^2   \nonumber \\
&+ 2[(\partial_0 \pi_i)(\partial_0 \pi_j)(\partial_i \pi_j) - (\partial_0 \pi_i)^2 (\partial_j \pi_j)] + 2[(\partial_j \pi_k)^2 (\partial_i \pi_i) - (\partial_k \pi_i)(\partial_k \pi_j)(\partial_i \pi_j)]  \nonumber \\
&+ (\partial_0 \pi_i)(\partial_0 \pi_j)(\partial_k \pi_i)(\partial_k \pi_j) - (\partial_0 \pi_i)^2 (\partial_j \pi_k)^2.
\end{align}
With gravity turned on in unitary gauge, we have:
\begin{align}
M^{a b} &= g^{a b} = \eta^{a b} - h^{a b} + \eta_{\mu \nu} h^{\mu a} h^{\nu b} + o(h^3)\,, \\
x &= D - h_{i i} - h_{0 i}^2 + h_{i j}^2 + o(h^3) \,, \\
y &= 1 - h_{i i} - h_{i 0}^2 + \frac{1}{2} h_{i i}^2 + \frac{1}{2} h_{i j}^2 + o(h^3)\,, \label{detM} \\
\sqrt{-g} &= 1 -\frac{1}{2} h_{0 0} + \frac{1}{2} h_{i i} - \frac{1}{8} h_{0 0}^2 + \frac{1}{2} h_{0 i}^2 - \frac{1}{4} h_{i j}^2 + \frac{1}{8} h_{i i}^2 - \frac{1}{4} h_{0 0} h_{i i} + o(h^3) \,. \label{detg}
\end{align}

Here $x = \text{Det}_1 (M)$ and $y = \text{Det}_D (M)$.

\subsection{Perfect Fluid}

For the case of an ideal fluid, our Lagrangian will only be a function of y, the determinant of the deformation matrix. Therefore, the  mass term will come from:
\begin{align}
\mathcal{L}_m = \sqrt{-g} [f_0 + f_1 (y - 1) + \frac{1}{2} f_2 (y-1)^2]\,,
\end{align}
where $f_n$ is just the nth derivative of f with respect to y evaluated when $y = 1$. This expansion can be considered as an expansion about the restframe state. The coefficients have physical significance in that the restframe energy density $\rho_0$, the restframe pressure $P_0$, and the restframe bulk modulus $K_0$ are given by $\rho_0 = -f_0, P_0 = f_0 - 2 f_1$ and $K_0 = -f_0 (1 + \frac{2 f_2}{f_1})$, respectively. The coefficients to higher order in the expansion would amount to giving corrections to these quantities away from the restframe state but will be unimportant for our analysis. In unitary gauge, the perfect fluid Lagrangian coupled to gravity to quadratic order in h in any dimension D is:
\begin{align}
\mathcal{L}_m &= f_0 +(-\frac{1}{2} f_0) h_{0 0} + (\frac{1}{2} f_0 - f_1) h_{i i} \nonumber \\
&+ (- \frac{1}{8} f_0) h_{0 0}^2 + (\frac{1}{2} f_0 - f_1) h_{0 i}^2 - \frac{1}{2} (\frac{1}{2} f_0 - f_1) h_{i j}^2 \nonumber \\
&+ \frac{1}{2} (\frac{1}{4} f_0 + f_2)h_{i i}^2 -(\frac{1}{4} f_0 - \frac{1}{2} f_1) h_{0 0} h_{i i}. \label{fluidmassivegrav}
\end{align}
We therefore have the following mass terms:
\begin{align}
m_0^2 &= - \frac{1}{8} f_0 = \frac{1}{8} \rho_0\,, \nonumber \\
m_1^2 &= \frac{1}{2} f_0 - f_1 = \frac{1}{2} P_0\,, \nonumber \\
m_2^2 &= \frac{1}{2} f_0 - f_1 = \frac{1}{2} P_0\,, \nonumber \\
m_3^2 &= \frac{1}{4} f_0 + f_2 = \frac{1}{4} P_0 - \frac{K_0 (P_0 + \rho_0)}{4 \rho_0} \,, \nonumber \\
m_4^2 &= \frac{1}{4}f_0 - \frac{1}{2} f_1 = \frac{1}{4} P_0\,.
\end{align}
While it would seem we are in Case 1 ($m_0, m_1 \ne 0$) of Lorentz violating massive gravity, there is a subtlety. Due to the linear  and constant terms in our Lagrangian, the field $\psi = h_{0 0}$ is hidden inside a Lagrange multiplier  which puts a constraint on $\tau$ and $\sigma$, reducing the degrees of freedom in the scalar sector by 1. This is manifest at the fully nonlinear level in ADM formalism shown later on. We define:
\begin{align}
N = 1 - \frac{1}{2} \psi - \frac{1}{8} \psi^2 + \frac{1}{2} \partial_i v \partial^i v\,.
\end{align}
Once we do this and integrate out $v$, our Lagrangian becomes:
\begin{align}
\mathcal{L}_S = \frac{1}{2} \tau \Big( D (D-1) \partial_0^2 - (D-1)(D-2) \nabla^2 + D^2 m_3^2 - Dm_2^2 - \frac{(D-1)^2}{f_1} \partial_0^2 \nabla^2 \Big) \tau \nonumber \\ + \sigma \Big( \frac{1}{2} (m_3^2 - m_2^2)\nabla^2 \nabla^2 \Big) \sigma + \tau \Big( (D-1) \partial_0^2 \nabla^2 + (D m_3^2 - m_2^2)\nabla^2 \Big) \sigma \nonumber \\
+N(1+ \frac{1}{2} (D \tau + \nabla^2 \sigma))[f_0 - (D - 1) \nabla^2 \tau - f_1 (D \tau + \nabla^2 \sigma)]\,. 
\end{align}

Here we have separated the constraint into two terms the first of which is $\sqrt{\gamma}$ expanded to first order in $h_{ii}$ where $\gamma_{i j} = g_{i j}$ is the ADM spatial metric and the second term is the true constraint which the equation of motion forces to be zero. If we naively expanded the whole quantity to first order in $\tau$ and $\sigma$ we would not obtain the correct constraint to lowest order in $\tau$ and $\sigma$. The constraint on $\tau$ and $\sigma$ is thus:

\begin{align}
\nabla^2 \sigma = -\frac{(D-1)}{f_1} \nabla^2 \tau - D \tau + \frac{f_0}{f_1}.
\end{align}
Then, substituting back in, we get a Lagrangian entirely in terms of $\tau$:
\begin{align}
\mathcal{L}_S &= \dot{\tau} \Big( \frac{D (D-1)}{2} + \frac{2 (D-1)^2}{(-f_1)} (-\nabla^2) \Big) \dot{\tau} \nonumber \\
&- \tau \Big(\frac{(D-1)^2(m_2^2 - m_3^2)}{2 f_1^2} \nabla^2 \nabla^2 +[\frac{(D-1)^2 m_2^2}{(-f_1)} + \frac{(D-1)(D-2)}{2}] (-\nabla^2) + \frac{D(D-1)}{2} m_2^2 \Big) \tau\,. \label{scalarfluidLag}
\end{align}
Here we did not write the linear term in $\tau$, since for our purposes in can be shifted away  by 
a field redefinition; we also  dropped the vacuum constant as it does not affect much the 
degrees of freedom at momenta higher that the scale 
set by the background. As we can see, there is manifestly no ghost unlike in the usual Lorentz violating massive gravity case when $m_0, m_1 \ne 0$ (see Case 1 of Section II.C).

In what follows, we wish to consider questions of stability and dynamics in three scenarios of interest. First, we will look at the fluid theory in the absence of gravity. Second, we will analyze the special case in which f evaluated in unitary gauge has no linear term in the perturbation $h_{ii}$ and thus, when $f = f_0$ is an exact stable background. Finally, we analyze the case of a general function f with linear term but for which the decay time from $f = f_0$ to the true vacuum is large enough that we can ignore the linear term and just consider the quadratic fluctuations for a general function f. In all three scenarios, we will consider what consistency conditions of stability and subluminal wave propagation impose on the function $f$ and thus on the equation of state. It will turn out that these consistency conditions for the fluid coupled to the gravitational field add no new constraints on the equation of state that weren't already forced upon us in the absence of gravity.

\subsubsection*{Case 1} 

The Lagrangian for the fluid up to quadratic order in the $\pi^a$ fields in the absence of gravity is:
\begin{align}
\mathcal{L}_m &= -f_1 (\partial_0 \pi_i)^2 + (f_1 + 2 f_2)(\partial_i \pi_i)^2 \label{fluidnograv}\,.
\end{align} 
From this Lagrangian, we see that for the kinetic and potential terms to have the correct sign, we must have the following conditions satisfied:
\begin{align}
f_1 &\le 0 \Rightarrow P_0 \ge -\rho_0 \nonumber \\
f_1 + 2 f_2 &\le 0 \Rightarrow \rho_0 \ge K_0. \label{stability}
\end{align}
With these conditions met, we have one healthy propagating degree of freedom, the longitudinal mode of $\pi^a$. Furthermore, the wave speed of this mode (the speed of sound in the medium) is given by:
\begin{align}
c_S^2 = 1 + 2\frac{f_2}{f_1} \label{soundspeed}\,.
\end{align}
Therefore, to ensure subluminal propagation we impose the condition $0 \le c_s^2 \le 1$ which corresponds to a restriction on the f coefficients:
\begin{align}
-\frac{1}{2} \le \frac{f_2}{f_1} \le 0. \label{longsoundspeed}
\end{align}

or equivalently:

\begin{align}
0 \le K < \rho_0
\end{align}

\subsubsection*{Case 2}
Now if we turn on gravity and require the the restframe of the fluid to be the true vacuum of the theory, we are forced to impose the more restrictive condition on the equation of state that $f_1 = 0$. Since there is no time dependent quadratic term, this special type of fluid would have instant response to any disturbance in the density across the whole medium instead of some local disturbance giving rise to a propagating longitudinal sound wave. This condition indeed enforces the constraint on the equation of state  to be, $\rho_0 = - P_0$,  as is the case for dark energy. As for looking at the gravitational waves in unitary gauge, only $m_3$ is nonzero and we certainly do not have a propagating degree of freedom in either the vector or scalar sector. Such a realization of "dark energy" seems to be bizarre (and most likely unphysical) due to the above instantaneous interactions.  Hence, we will not pursue this case further.

\subsubsection*{Case 3}

Now if we turn on gravity, we can immediately see that we have a propagating transverse vector as well as a propagating scalar. At first, the fact that the graviton gains two dynamical degrees of freedom and particularly, gains a transverse mode might seem to defy expectations for what degrees of freedom should propagate since when gravity is turned off, we have only a longitudinal propagating mode in the fluid. However, the interaction of gravity with the fluid in fact turns on the extra transverse degree of freedom. Indeed, the transverse mode is hiding in the gravitational perturbation field all along. This can easily be seen introducing back the $\phi^a$ field from our Lagrangian in unitary gauge as a Stueckelberg field for the spatial diffeomorphisms:
\begin{align}
h_{0i} = \bar{h}_{0i} - \partial_0 \pi_i \,,\nonumber \\
h_{i j} = \bar{h}_{i j} - \partial_i \pi_j - \partial_j \pi_i\,.
\end{align}
When we plug these fields into the Lagrangian \eqref{fluidmassivegrav}, and just look at the quadratic part of the Lagrangian solely dependent on $\pi^a$, we get:
\begin{align}
\mathcal{L}_{\pi} = -f_1 (\partial_0 \pi_i)^2 + (f_1 + \frac{1}{2} f_2)(\partial_i \pi_i)^2 + f_1 (\partial_i \pi_j)^2.
\end{align}
Indeed, we see that the Lagrangian for the fluid in the absence of gravity is reproduced \eqref{fluidnograv} with the addition of the last term which is the transverse mode that was hiding in the metric perturbation field. 

In the vector sector, we have a transverse vector propagating with dispersion relation: 

\begin{align}
\omega^2 = \frac{P_0}{P_0 + \rho_0} k^2 + \frac{1}{2} P_0
\end{align}

This is clearly subluminal and no constraints need to be imposed.

 In the scalar sector, our Lagrangian for $\tau$ and the dispersion relation are very similar to those of Case 5. Our positive energy conditions in this case are:

\begin{align}
\frac{2 (D-1)^2}{(-f_1)} \ge 0 \label{poskinfluid} \\
m_2^2 \ge m_3^2 \label{pospotfluid}
\end{align}

These are already guaranteed to hold by the consistency conditions in the absence of gravity. Our dispersion relation for the propagating gravitational scalar mode is (see \eqref{dispersion}):
\begin{align}
\omega^2 = \frac{\frac{(D-1)^2 (m_2^2 - m_3^2)}{2 f_1^2} k^4 + [\frac{(D-1)^2 m_2^2}{(-f_1)} +\frac{(D-1)(D-2)}{2}]k^2 + \frac{D(D-1)}{2} m_2^2}{\frac{D (D-1)}{2} + \frac{2 (D-1)^2}{(-f_1)} k^2}.
\end{align}

Therefore, we can see that subluminal propagation condition is:

\begin{align}
\frac{m_2^2-m_3^2}{4(-f_1)} \le 0 \label{sublumfluid}
\end{align}

If we translate this into a condition on the physical constants, the constraint is that:

\begin{align}
\frac{1}{8} \Big( \frac{P_0}{P_0 + \rho_0} + \frac{K_0}{\rho_0} \Big) \le 1.
\end{align}

This condition is trivially satisfied however since all the quantities are positive and $K_0 \le \rho_0$ so no new constraint is placed on the equation of state. 

\subsection{Elastic Solid}

For the case of an elastic solid, the Lagrangian can depend on all the rotational invariants of $M^{a b}$ which in the case of 2 spatial dimensions, means that our function f depends on both x and y. We will be expanding f about $f(2,1)$ to quadratic order in h. First we introduce some simplifying notation for the first few partial derivatives of f at $x = 2, y = 1$:
\begin{align}
f_{0 0} &= f(2, 1)\,, & f_{1 1} &= \partial_x \partial_y f (2, 1)\,, \nonumber \\
f_{1 0} &= \partial_x f (2, 1)\,, & f_{2 0} &= \partial_x^2 f (2, 1)\,, \nonumber \\
f_{0 1} &= \partial_y f (2, 1)\,, & f_{0 2} &= \partial_y^2 f (2, 1)\,.
\end{align}
With this notation, the mass term in the Lagrangian becomes:
\begin{align}
\mathcal{L}_m &= f_{0 0} + (- \frac{1}{2} f_{0 0}) h_{0 0} + (\frac{1}{2} f_{0 0} - f_{0 1} - f_{1 0}) h_{i i} \nonumber \\
&(- \frac{1}{8} f_{0 0}) h_{0 0}^2 + (-\frac{1}{2} f_{0 0} - f_{1 0} - f_{0 1}) h_{0 i}^2 - \frac{1}{2} (\frac{1}{2} f_{0 0} - 2 f_{1 0} - f_{0 1}) h_{i j}^2 \nonumber \\
&+ \frac{1}{2}(\frac{1}{4} f_{0 0} - f_{1 0} + f_{2 0} + f_{0 2} + 2 f_{1 1}) h_{i i}^2 - (\frac{1}{4} f_{0 0} - \frac{1}{2} f_{1 0} - \frac{1}{2} f_{0 1}) h_{00} h_{ii}\,.
\end{align}
Therefore, the mass parameters are expressed as:
\begin{align}
m_0^2 &= - \frac{1}{8} f_{0 0}\,, \nonumber \\
m_1^2 &= \frac{1}{2} f_{0 0} - f_{1 0} - f_{0 1} \,,\nonumber \\
m_2^2 &= \frac{1}{2} f_{0 0} - 2 f_{1 0} - f_{0 1}\,, \nonumber \\
m_3^2 &= \frac{1}{4} f_{0 0} - f_{1 0} + f_{2 0} + f_{0 2} + 2 f_{1 1} \,, \nonumber \\
m_4^2 &= \frac{1}{4} f_{0 0} - \frac{1}{2} f_{1 0} - \frac{1}{2} f_{0 1} \,.
\end{align}
Some of these coefficients have clear physical meaning. If we approximate our elastic isotropic solid as a Hooke's law solid in which the energy density only depends quadratically on the strain matrix $s_{a b} \equiv \frac{1}{2} (\delta_{a b} - M_{a b})$, then the matter Lagrangian function f can be written as:
\begin{align}
f = -\rho_0\sqrt{y}-\frac{1}{2} \sqrt{y} ( K s^2 + 2 \mu r_{a b} r_{a b})\,,
\end{align}
where $s = \delta^{a b} s_{a b}$ and $r_{a b}$ is the traceless part of $s_{a b}$ \cite{Soper2008}. This is a very good approximation to most isotropic elastic solids which are close to their equilibrium state. Here, the energy density is $\rho_0 = - f_{00}$, K is the bulk modulus and $\mu$ is the shear modulus. The other coefficients can then be computed as:
\begin{align}
f_{10} &= -\frac{1}{2} \mu \,, \nonumber \\
f_{01} &= -\frac{1}{2} (\rho_0-\mu) \,, \nonumber \\
f_{20} &= -\frac{1}{4} (K + \mu)\,, \nonumber \\
f_{11} &= -\frac{1}{4} \mu \,, \nonumber \\
f_{02} &= \frac{1}{4} \rho_0 + \frac{1}{2} \mu \,.
\end{align} 

and in the Hooke's Law approximation the masses are explicitly given in terms of the physical constants by:
\begin{align}
m_0^2 &= \frac{1}{8} \rho_0 \,, \nonumber \\
m_1^2 &= 0 \,,\nonumber \\
m_2^2 &= \frac{\mu}{2}  \,, \nonumber \\
m_3^2 &= \frac{\mu-K}{4}  \,, \nonumber \\
m_4^2 &= 0  \,. \label{solidmasses}
\end{align}

From these masses, it might naively seem like we are in Case 2 from Section 1 in which there were no dynamic degrees of freedom since $m_0 \ne 0$ and $m_1 = 0$. However, for the same reason as with the perfect fluid case, $\psi$ is really contained inside a Lagrange multiplier, the lapse in ADM formalism, and this effects the true dynamics of the scalar sector. Therefore, the Lagrangian for the scalar sector is the same as it was for the perfect fluid \eqref{scalarfluidLag} except with the masses given in terms of the physical constants by \eqref{solidmasses} and $f_1$ replaced with $f_{0 1} + f_{1 0}$. 

We consider the three cases for the elastic solid. In the first case, we look at the theory in the absence of gravity and see what the constraints from ensuring a positive Hamiltonian and subluminal sound speeds impose on the derivatives of f. In the second case, we impose extra conditions on f to make the restframe of the solid the true vacuum and see what the physical implications are for the solid. In the third case, we consider the dynamics of the theory for a general f assuming the decay time to the true vacuum is large enough that the linear terms in the Lagrangian can be ignored. By requiring consistency of the gravitational theory, additional constraints on $f$ are imposed.

\subsubsection*{Case 1}

In the absence of gravity, our Lagrangian for the $\pi^a$ fields to quadratic order becomes:
\begin{align}
\mathcal{L}_m &= -(f_{10}+f_{01})(\partial_0 \pi_i)^2 + f_{10} (\partial_i \pi_j)^2 + (f_{01} + 4 f_{11} + 2 f_{20} +2 f_{02})(\partial_i \pi_i)^2 \,.
\end{align}

To ensure correct signs for the kinetic and potential terms, we must impose the following conditions on $f$:
\begin{align}
f_{10}+f_{01} &\le 0 \Rightarrow \rho_0 \ge 0. \nonumber \\
f_{10} &\le 0 \Rightarrow \mu \ge 0. \nonumber \\
f_{01} + 4 f_{11} + 2 f_{20} +2 f_{02} &\le 0 \Rightarrow K \ge 0.
\end{align}
These stability conditions naturally imply a non-negative energy density, shear modulus, and bulk modulus as we would expect. We can see from this Lagrangian that we get two propagating degrees of freedom: 1 longitudinal mode propagating at speed $c_S$ (the longitudinal sound speed) and 1 transverse mode propagating at speed $c_T$ (the transverse sound speed) as we would expect for a solid. These speeds are:
\begin{align}
c_S^2 &= \frac{(f_{10}+f_{01} + 4 f_{11} + 2 f_{20} +2 f_{02})}{f_{10}+f_{01}} = \frac{K+\mu}{\rho_0}\,, \nonumber \\
c_T^2 &= \frac{f_{10}}{f_{10}+f_{01}} = \frac{\mu}{\rho_0} \,.
\end{align}
Enforcing the condition that $0 \le c_S^2 \le 1$ gives us the additional constraint:

\begin{align}
K + \mu \le \rho_0 \nonumber \\ \label{nogravsolidconstraint}
\end{align}

The condition that $0 \le c_T^2 \le 1$ adds no new information. If our solid is furthermore conformally symmetric, then $c_S$ and $c_T$ are no longer independent but satisfy \cite{Esposito2017}:

\begin{align}
c_S^2 = \frac{1}{2} + c_T^2
\end{align}
 
 in 2 dimensions. This interestingly gives an exact relationship between bulk modulus and energy density in a 2D solid: $K = \frac{\rho_0}{2}$ and by virtue of \eqref{nogravsolidconstraint}, imposes that $\mu \le K$. However, we will not assume conformal symmetry moving forward. 

\subsubsection*{Case 2}

Next, it is interesting to turn on gravity and impose an extra condition to ensure that the restframe of the 2D solid is the true vacuum of the theory. For this to be true, we must then impose an extra condition on f that $f_{10}+f_{01} = 0$. When we impose this special condition, we eliminate the quadratic term with time derivatives therefore eliminating the propagation of sound waves. We get a peculiar solid in which local compressions and shears are felt instantly throughout the entire medium making the solid ``infinitely rigid'' in a sense. However, the physical meaning of such a condition is that the rest energy is zero which is unphysical. In unitary gauge, this condition forces all masses to be zero except $m_2$ and $m_3$ where again, we can see that there are no propagating degrees of freedom. We therefore drop this case and move on to the more general one.

\subsubsection*{Case 3}

Leaving f general, we again have a propagating transverse vector mode with dispersion relation:

\begin{align}
\omega^2 = \frac{\mu}{\rho_0} k^2 + \frac{\mu}{2}
\end{align}

As for the scalar mode, we have the same positive energy conditions \eqref{poskinfluid}-\eqref{pospotfluid} and subluminal propagation condition \eqref{sublumfluid} as we did for the fluid case save for the different values of the masses and the replacement of $f_{1 0} + f_{0 1}$ instead of $f_1$. The positive energy conditions are equivalent to requiring $\rho_0 \ge 0$ and $\mu + K \ge 0$, which are both guaranteed from the stability of the solid in the absence of gravity. The subluminal propagation condition is equivalent to requiring:

\begin{align}
\frac{1}{8} \frac{\mu + K}{\rho_0} \le 1
\end{align}
 
 This actually adds no new condition since we already know that $\mu + K \le \rho_0$ from the analysis of the solid in the absence of gravity. Therefore, stability and subluminal wave propagation of phonons at large momenta in the presence of gravitational interactions impose no additional constraints on the equation of state of the 2D solid. 

\subsection{Hamiltonian Analysis}

Finally, we would like to analyze the degrees of freedom of our system for the full nonlinear Lagrangian $\mathcal{L} = \mathcal{L}_G + \mathcal{L}_m$ where $\mathcal{L}_G$, the gravitational Lagrangian and $\mathcal{L}_m$, the matter Lagrangian are defined in terms of ADM variables as:
 \begin{align}
\mathcal{L}_G &= \sqrt{\gamma} N [^{(D)} R + K_{i j} K^{i j} - K^2]\,, \\
\mathcal{L}_m &= \sqrt{\gamma} N f(\text{Det}_1 (M),\dots, \text{Det}_D (M))\,.
\end{align}
Here, $^{(D)}R$ is the intrinsic scalar curvature of the spatial D manifold and $K_{i j}$ is the extrinsic curvature tensor given by the formula \cite{wald1984}:
\begin{align}
K_{i j} = \frac{1}{2} N^{-1} [\dot{\gamma}_{i j} - D_i N_j - D_j N_i]\,,
\end{align}
where all contractions are with respect to the spatial metric $\gamma_{i j}$,  and $D_i$ is the covariant derivative with respect to the spatial metric. The deformation matrix in unitary gauge is:

\begin{align}
M^{i j} = g^{i j} = \gamma^{i j} - n^i n^j
\end{align}

where $n_i \equiv \frac{N_i}{N}$. Therefore all invariants of $M$ will only depend on $\gamma_{i j}$ and $n_i$ but not $N$. Defining the canonical momenta $\pi_{i j}$ in the usual way as, $\pi_{i j} = \frac{\delta \mathcal{L}_G}{\delta \dot{\gamma}^{i j}}$, we get the following overall Hamiltonian:
\begin{align}
\mathcal{H} = \sqrt{\gamma} N \Big[ ^{(D)}R + \gamma^{-1} (\frac{1}{2} \pi^2 - \pi^{i j} \pi_{i j}) + 2 n_j D_i (\gamma^{-\frac{1}{2}} \pi^{i j})- f(x, y)  \Big]\,.
\end{align}
Since the lapse $N$,  and rescaled shift,  $n_i \equiv \frac{N_i}{N}$, appear in the Hamiltonian without time derivatives, they are not dynamical variables and therefore, their equations of motion give us constraints. Taking a variation of $\mathcal{H}$ with respect to N gives us the constraint that everything inside the large brackets is equal to zero. Taking the variation of $\mathcal{H}$ with respect to the rescaled shift function gives us the the equations of motion for $n_i$ in terms of only $n_i, \gamma_{i j},$ and $\pi_{i j}$. This implicitly gives $n_i$ in terms of $\gamma_{i j}, \pi_{i j}$. Therefore, we can integrate out $n^i$ making the only variables of consequence the D-dimensional spatial metric and the conjugate momenta. Then $N$ acts as an overall Lagrange multiplier which gives a constraint on $\gamma_{i j}, \pi_{i j}$. Therefore, the number of degrees of freedom can be counted as the number of degrees of freedom in $\gamma_{i j}$ minus one ($\frac{D(D+1)}{2}-1$) which exactly corresponds to 1 transverse, traceless tensor, 1 transverse vector and 1 scalar degree of freedom) and therefore, there is no extra scalar ghost as we discussed previously in the fluid and solid cases. 

\section{\label{sec:level1} 3D Massive Gravity as a Massive Chern-Simons Theory}

While our formalism in the previous section covered a continuous medium coupled to gravity in any dimension under fairly general conditions, it is interesting to focus in on the $D = 2$ case in which gravity has a dual description in terms of a Chern-Simons gauge field and study the dynamics of such a theory in the gauge field language. We do not consider a general Lorentz-violating mass term for the gauge field however but only the special case of a mass term which corresponds to the Fierz-Pauli term in the metric language. We find that this theory has two degrees of freedom and no ghosts as we expect from the well-known results for Fierz-Pauli theory in (2+1)-dimensions.

\subsection{Chern-Simons Gravity}

In (2+1)-dimensions, gravity is special in that it can be shown to be equivalent (at least perturbatively) 
to a Chern-Simons theory with gauge group being the symmetry group of the vacuum of the theory \cite{Achucarro1986, WITTEN1986}. In the first order formalism, we can define an orthonormal one form triad that forms a basis for the space of one forms $( e^a = e^a_\mu dx^{\mu}, a = 0, 1, 2 )$ on the manifold. The orthonormality condition gives us that:
\begin{align}
g_{\mu \nu} = \eta_{a b} e^a_{\mu} e^b_{\nu}.
\end{align} 
Along with these, we have the spin-connection $\omega_{\mu}^{a b}$. In terms of the spin connection, we define the curvature two form as $\tensor{R}{^a_b}$:
\begin{align}
\tensor{R}{^a_b} = d\tensor{\omega}{^a_b} + \tensor{\omega}{^a_c} \wedge \tensor{\omega}{^c_b}\,.
\end{align}
And with this we can define the Einstein-Hilbert action plus cosmological constant as:
\begin{align}
S = \frac{1}{8 \pi G} \int \epsilon_{a b c} R^{a b} \wedge e^c + \frac{1}{6} \lambda e^a \wedge e^b \wedge e^c \epsilon_{abc}\,.
\end{align}
We define a vector valued one form from our spin-connection as
\begin{align}
\omega^a \equiv \frac{1}{2} \epsilon^{a b c} \omega_{b c}\,,
\end{align}
 and then with that, we can define two gauge fields:
 \begin{align}
 A^{\pm a} \equiv \omega^a \pm \frac{1}{l} e^a\,,
 \end{align}
 where $l$ is the AdS radius ($\lambda = -\frac{1}{l^2}$). Both of our gauge fields transform under gauge group SL(2, $\mathcal{R}$). The Einstein-Hilbert action (101) in terms of the gauge fields is then:
 \begin{align}
 S = \frac{l}{32 \pi G} \Big( S_{CS} [A^{+ a}] - S_{CS} [A^{- a}] \Big)\,,
 \end{align}
 where, $S_{CS}$ denotes the Chern-Simons action:
\begin{align}
S_{CS}[A^a] = \int A^a \wedge dA^a + \frac{2}{3} \epsilon_{a b c} A^a \wedge A^b \wedge A^c \,.
\end{align}
If we define an orthonormal basis in AdS, $e^a_{(0)}$, and the associated spin connection $\omega^a_{(0)}$, then we can define the gauge fields as background fields plus perturbations:
 \begin{align} \label{gaugeperturb}
 A^{\pm a} = \mathcal{A}^{\pm a} \pm a^{\pm a}\,,
 \end{align}
 where
 \begin{align}
  \mathcal{A}^{\pm a}  = \omega^a_{(0)} \pm \frac{1}{l} e^a_{(0)}\,.
 \end{align}
 The Chern-Simons gravity action written out in full is:
 \begin{align}
 &\frac{l}{32 \pi G} \int \Big[ A^{+ a} \wedge dA^{+ a} + \frac{2}{3} \epsilon_{a b c} A^{+ a} \wedge A^{+ b} \wedge A^{+ c}\Big] - \Big[ A^{- a} \wedge dA^{- a} + \frac{2}{3} \epsilon_{a b c} A^{- a} \wedge A^{- b} \wedge A^{- c}\Big]  \\
 = &\frac{l}{16 \pi G}\int d^3 x \sqrt{-g} \epsilon^{\mu \nu \rho}  \Big[A^{+ a}_\mu \partial_\nu A^{+ a}_\rho + \frac{1}{3} \epsilon_{a b c} A^{+ a}_\mu A^{+ b}_\nu A^{+ c}_\rho - A^{- a}_\mu \partial_\nu A^{- a}_\rho - \frac{1}{3} \epsilon_{a b c} A^{- a}_\mu A^{- b}_\nu A^{- c}_\rho \Big]\,. 
 \end{align}
 If we decompose the gauge fields as in \eqref{gaugeperturb}, then we will have terms in the Lagrangian of orders zero, one, two, and three in the perturbations. The zero order is background which we can ignore in dealing with the gauge dynamics. The third order we can also ignore as we are only interested in the linearized equations of motion for the perturbations. The first order terms  vanish  due to the equations of motion. Thus, we have the following quadratic Lagrangian for the perturbations:
 
 \begin{align}
 \frac{l}{16 \pi G}\int d^3 x \sqrt{-g^{(0)}} \epsilon^{\mu \nu \rho}  \Big[ a^{+ a}_\mu \partial_\nu a^{+ a}_\rho +\epsilon_{a b c} a^{+ a}_\mu a^{+ b}_\nu \mathcal{A}^{+ c}_\rho \Big] - \epsilon^{\mu \nu \rho} \Big[ a^{- a}_\mu \partial_\nu a^{- a}_\rho + \epsilon_{a b c} a^{- a}_\mu a^{- b}_\nu \mathcal{A}^{- c}_\rho \Big]\,.
 \end{align}
 We now wish to add a Fierz-Pauli mass term to the Lagrangian and understand the dynamics and degrees of freedom in terms of the gauge field perturbations $a^{\pm a}_\mu$ instead of the usual metric perturbations $h_{\mu \nu}$. The Fierz-Pauli mass term is:
 \begin{align}
 \mathcal{L}_{FP} = -\frac{1}{4} m^2 \sqrt{-g^{(0)}} g^{(0) \mu \alpha} g^{(0) \nu \beta} (h_{\mu \nu} h_{\alpha \beta} - h_{\mu \alpha} h_{\nu \beta})\,,
 \end{align}
 where $g^{(0)}_{\mu \nu}$ denotes the AdS$_3$ metric and $h_{\mu \nu}$ is the perturbation from that metric. We know $h_{\mu \nu}$ to linear order in the gauge field perturbations is:
 \begin{align}
 h_{\mu \nu} = \frac{l}{2} \eta_{a b} \Big[ e^a_{(0) \mu} (a^{+ b}_\nu + a^{- b}_\nu) + e^b_{(0) \nu} (a^{+ a}_\mu + a^{- a}_\mu)  \Big]\,.
 \end{align}
Then the full Lagrangian becomes:
\begin{align}
\frac{l}{16 \pi G}&\int d^3 x \sqrt{-g^{(0)}} \Big[ \epsilon^{\mu \nu \rho}  \Big[a^{+ a}_\mu \partial_\nu a^{+ a}_\rho +\epsilon_{a b c} a^{+ a}_\mu a^{+ b}_\nu \mathcal{A}^{+ c}_\rho - a^{- a}_\mu \partial_\nu a^{- a}_\rho - \epsilon_{a b c} a^{- a}_\mu a^{- b}_\nu \mathcal{A}^{- c}_\rho \Big]  \nonumber \\
&-\frac{m^2 l}{8} \Big[ \Big( (a^{+ a}_\mu + a^{- a}_\mu)(a^{+ \mu}_a + a^{- \mu}_a) + (a^{+ a}_\mu + a^{- a}_\mu)(a^{+ b}_\nu + a^{- b}_\nu) e^\mu_{(0) b} e^\nu_{(0) a} \Big) \nonumber \\
&-2\Big( e^\mu_{(0) a} (a^{+ a}_\mu + a^{- a}_\mu) \Big)^2 \Big] \Big]
\end{align}
In the limit as we take the AdS radius radius $l$ to infinity while keeping fixed the quantity $M \equiv \frac{m^2 l}{4}$, we get the flat spacetime limit of this theory while preserving the degrees of freedom:
\begin{align}
\mathcal{L} = &\eta_{a b} \epsilon^{\mu \nu \rho} a^{+ a}_\mu \partial_\nu a^{+ a}_\rho - \eta_{a b} \epsilon^{\mu \nu \rho} a^{- a}_\mu \partial_\nu a^{- a}_\rho \nonumber \\
&-\frac{M}{2} (a^+ + a^-)^a_\mu (a^+ + a^-)^b_\nu (\eta_{a b} \eta^{\mu \nu} + \delta^\mu_b \delta^\nu_a - 2 \delta^\mu_a \delta^\nu_b)
\end{align}
We would like to figure out how many degrees of freedom are propagated in the flat space limit of the AdS Chern-Simons massive gravity. Let us start with a few simpler theories and work our way up.

\subsection{Massive Abelian Chern-Simons Theory}

First we consider the following Chern-Simons Abelian gauge theory  with a mass term:
\begin{align}
\mathcal{L} = \epsilon^{\mu \nu \rho} a_\mu \partial_\rho a_\nu - m \eta^{\mu \nu} a_\mu a_\nu.
\end{align}
Such a theory can arise in condensed matter systems and the development of such a mass term in a Chern-Simons gauge theory is known as the Abelian Chern-Simons-Higgs mechanism \cite{Dunne1999}. In this theory there is one massive degree of freedom. There are several ways to see this. First of all, the Causal propagator for this theory is:
\begin{align}
\Delta_{\mu \nu} = \frac{-1}{m} \frac{\epsilon_{\mu \nu \lambda} p^{\lambda} + i (\eta_{\mu \nu} + \frac{p_\mu p_\nu}{m^2})}{p^2 + m^2-i\epsilon}.
\end{align}
which shows that there is a single pole at $p^2 = -m^2$ and thus a particle of mass m in the theory. We can see that there is one degree of freedom in the theory by introducing a Stueckelberg field as follows:
\begin{align}
\bar{a}_\mu = a_\mu + \frac{1}{\sqrt{m}} \partial_\mu \phi
\end{align}
with the gauge symmetry:
\begin{align}
a_\mu &\rightarrow a_\mu + \frac{1}{\sqrt{m}} \partial_\mu \gamma\,, \nonumber \\
\phi &\rightarrow \phi - \gamma\,.
\end{align}
In the decoupling limit as $m \rightarrow 0$, we get a single propagating massless scalar degree of freedom and 
a non-propagating Chern-Simons field. 

Finally, if we decompose the gauge field $a_\mu$ into rotationally irreducible representations as follows:
\begin{align}
a_0 &= \phi \,, \nonumber \\
a_i &= \partial_i \sigma + s_i\,,
\end{align}
where $s_i$ is transverse, we can see that $\phi$ and $\sigma$ can be integrated out and we get:
 \begin{align}
 \mathcal{L} = \frac{1}{m} s_i \Big[ -\partial_0^2 + \nabla^2 -m^2 \Big] s_i\,,
 \end{align}
 which shows that our degree of freedom is a massive transverse vector field with the usual dispersion relation. This is in agreement with the Stueckelberg analysis because in 3 dimensions, there is a duality between scalars and transverse vectors in that any transverse vector field $s_i$ can always be written as $s_i = \epsilon_{i j} \partial_j \theta$ for some scalar field $\theta$. We can also see from this resulting Lagrangian for $s_i$ that the energy is positive definite.
 
\subsection{Massive $SL(2,R)$ Chern-Simons Theory}

The next theory we would like to analyze is similar to that which arises from Chern-Simons gravity except that we only have one $SL(2, R)$ gauge field:
\begin{align}\label{CSmassGauge}
\mathcal{L} = \epsilon^{\mu \nu \rho} \eta_{ab} a^a_\mu \partial_\rho a^b_\nu - \frac{m}{2} a^a_\mu a^b_\nu (\eta_{a b} \eta^{\mu \nu} + \delta^\mu_b \delta^\nu_a - 2 \delta^\mu_a \delta^\nu_b)\,.
\end{align}
The first interesting thing to notice about this theory is that it has an unusual gauge symmetry. The action is invariant under the following transformation:
\begin{align} \label{gaugesym}
a^a_\mu \rightarrow a^a_\mu +\tensor{\epsilon}{^a_\mu_\lambda} \partial^\lambda \phi + \frac{1}{m} \partial^a \partial_\mu \phi\,,
\end{align}
where $\phi$ is an arbitrary function. One can understand the origin of this gauge symmetry by seeing that the tensor structure of the mass term is similar but not identical to $\epsilon_{a b c} \epsilon^{\mu \nu \lambda} \delta^c_\lambda$ which accompanies the quadratic expansion about the background gauge field of the cubic part of the non-Abelian SL(2, $\mathcal{R}$) Chern-Simon's action. The difference between the  two tensor structures is due to 
the Fierz-Pauli mass term. Thus, the quadratic mass term in \eqref{CSmassGauge}  can be thought to originate
from two sources, from the cubic term in the non-Abelian CS theory calculated on a  background,   and 
the quadratic FP mass term.  It is interesting that such a theory has gauge invariance \eqref{gaugesym}. 
If one forgets about the factor of $\frac{1}{m}$ and replaces $\partial_a \phi$ with $\pi_a$, then \eqref{gaugesym} is identical to an infinitesimal gauge transformation parametrized by $\pi_a$. 
 Because of this gauge symmetry, the propagator is ill-defined without a gauge-fixing term in the Lagrangian. However, we are interested here in determining the degrees of freedom for this theory and for that purpose, we can introduce the following Stueckelberg field:
\begin{align}
\bar{a}^a_\mu = a^a_\mu + \frac{1}{\sqrt{m}} \partial_\mu V^a\,,
\end{align}
for which we have the gauge symmetry:
\begin{align}
a^a_\mu &\rightarrow a^a_\mu + \frac{1}{\sqrt{m}} \partial_\mu W^a \nonumber \\
V^a &\rightarrow V^a - W^a.
\end{align}
Upon taking the decoupling limit, our Lagrangian is then:
\begin{align}
\mathcal{L} = &\epsilon^{\mu \nu \rho} a^a_\mu \partial_\rho a^b_\nu - \frac{1}{2} \Big( (\partial_\mu V_\nu)^2 - (\partial_\lambda V^\lambda)^2 \Big)\,,
\end{align}
and we see that we have 1 propagating vector degree of freedom living inside  $V_\mu$. 

We can also determine the degrees of freedom using a decomposition similar to that for the Abelian case. We will start with our original Lagrangian and then decompose $a^a_\mu$ in the following way:
\begin{align}
a^0_0 &= \pi\,, \nonumber \\
a^i_0 &= \partial^i r + w^i \,, \nonumber \\
a^0_i &= \partial_i \phi + s_i \,, \nonumber \\
a^i_j &= \epsilon_{j i} \lambda + \partial_j v_i + \partial_i v_j + \partial_j \partial_i \sigma + \delta_{j i} \tau\,.
\end{align}
Here $w_i, s_i, v_i$ are all transverse vectors. It turns out that $\pi$ acts a Lagrange multiplier and all other field except $v_i$ we can integrate out. When we do this, we get the following Lagrangian:
 \begin{align}
 \mathcal{L} = v_i \Big[ -\partial_0^2 + \nabla^2 -m^2 \Big] \Big( \frac{-2 m \nabla^2}{-\nabla^2 + m^2} \Big) v_i
 \end{align}
If we just renormalize $v_i$ in the following way:
\begin{align}
v_i \rightarrow \sqrt{\frac{-2 m \nabla^2}{-\nabla^2 + m^2}} v_i\,,
\end{align}
we see then that we have one massive vector degree of freedom with the usual 
dispersion relation and in the massless limit, this agrees with our Stueckelberg analysis.

\subsection{Massive Chern-Simons Gravity}

Finally, we need to look at the Lagrangian we ended with in section 4.1:
\begin{align}
\mathcal{L} = &\eta_{a b} \epsilon^{\mu \nu \rho} a^a_\mu \partial_\nu a^b_\rho - \eta_{a b} \epsilon^{\mu \nu \rho} b^a_\mu \partial_\nu b^b_\rho \nonumber \\
&-\frac{m}{2} (a + b)^a_\mu (a + b)^b_\nu (\eta_{a b} \eta^{\mu \nu} + \delta^\mu_b \delta^\nu_a - 2 \delta^\mu_a \delta^\nu_b)\,,
\end{align}
where we have just renamed $a = a^+, b = a^-$ and $m = M$. Since this theory just came from Fierz-Pauli linearized gravity in 3 dimensions, it should propagate 2 degrees of freedom (1 transverse vector mode and 1 scalar mode). To see this, we will carry out a Stueckelberg analysis in the following way:
\begin{align}
\bar{a}^a_\mu &= a^a_\mu + \frac{1}{\sqrt{m}} \partial_\mu V^a\,, \nonumber \\
\bar{b}^a_\mu &= b^a_\mu + \frac{1}{\sqrt{m}} \partial_\mu W^a\,,
\end{align}
for which we have the gauge symmetries:
\begin{align}
a^a_\mu &\rightarrow a^a_\mu + \frac{1}{\sqrt{m}} \partial_\mu C^a\,, \qquad V^a \rightarrow V^a - C^a\,, \nonumber \\
b^a_\mu &\rightarrow b^a_\mu + \frac{1}{\sqrt{m}} \partial_\mu N^a \,, \qquad W^a \rightarrow W^a - N^a\,.
\end{align}
We can then introduce two additional Stueckelberg scalars in the following way patterned after the special gauge symmetry discussed in the previous section:
\begin{align}
\bar{a}^a_\mu &= a^a_\mu + \tensor{\epsilon}{^a_\mu_\lambda} \partial^\lambda \pi\,, \qquad \bar{V}^a = V^a + \frac{1}{\sqrt{m}} \partial^a \pi\,, \nonumber \\
\bar{b}^a_\mu &= b^a_\mu + \tensor{\epsilon}{^a_\mu_\lambda} \partial^\lambda \phi \,, \qquad \bar{W}^a = W^a - \frac{1}{\sqrt{m}} \partial^b \phi \,.
\end{align}
With these fields, we can take the decoupling limit $m \rightarrow 0$ of the Lagrangian and get:
\begin{align}
\mathcal{L} = &\eta_{a b} \epsilon^{\mu \nu \rho} a^a_\mu \partial_\nu a^b_\rho - \eta_{a b} \epsilon^{\mu \nu \rho} b^a_\mu \partial_\nu b^b_\rho \nonumber \\
&- 2 ( -\partial^\nu \partial_b \pi + \box \pi \delta^\nu_b) b^b_\nu + 2 ( -\partial^\nu \partial_b \phi + \box \phi \delta^\nu_b) a^b_\nu \nonumber \\
&- \frac{1}{2} \Big( (\partial_\mu (V+W)_\nu)^2 - (\partial_\lambda (V+W)^\lambda)^2 \Big)\,.
\end{align}
Now one  can see one vector degree of freedom with the usual Maxwell kinetic term,  but the scalars are coupled to the gauge fields and so we must make some field redefinitions to recover the kinetic term for the scalars. We make the following field redefinitions of the gauge fields:
\begin{align}
a^a_\mu &= \bar{a}^a_\mu + \tensor{\epsilon}{^a_\mu_\lambda} \partial^\lambda \phi - \frac{1}{16} \delta^a_\mu (\pi + \phi)\,, \nonumber \\
b^a_\mu &= \bar{b}^a_\mu + \tensor{\epsilon}{^a_\mu_\lambda} \partial^\lambda \pi + \frac{1}{16} \delta^a_\mu (\pi + \phi)\,. 
\end{align}
Under these field redefinitions, we are able to cancel the old couplings of $\pi$ and $\phi$ to the gauge fields and get a kinetic term:
\begin{align}
\mathcal{L} = &\eta_{a b} \epsilon^{\mu \nu \rho} \bar{a}^a_\mu \partial_\nu \bar{a}^b_\rho - \eta_{a b} \epsilon^{\mu \nu \rho} \bar{b}^a_\mu \partial_\nu \bar{b}^b_\rho - \frac{1}{8} (\phi + \pi) \tensor{\epsilon}{_b^\nu^\rho} \partial_\rho (\bar{a} + \bar{b})^b_\nu \nonumber \\
&-\frac{1}{2} \partial_\mu (\phi + \pi) \partial^\mu (\phi + \pi) -\frac{1}{2} \Big( (\partial_\mu (V+W)_\nu)^2 - (\partial_\lambda (V+W)^\lambda)^2 \Big)\,.
\end{align}
There is still a coupling of the scalars to the gauge fields which could potentially be a problem. However, it turns out to not be an issue at all. To show this most easily, let us make the following field redefinitions:
\begin{align}
c^a_\mu = \frac{1}{2} (\bar{a} + \bar{b})^a_\mu\,, \qquad d^a_\mu = \frac{1}{2} (\bar{a} - \bar{b})^a_\mu\,.
\end{align}
Then  our Lagrangian reads:
\begin{align}
\mathcal{L} = &\eta_{a b} \epsilon^{\mu \nu \rho} d^a_\mu \partial_\rho c^b_\nu - \frac{1}{4} (\phi + \pi) \tensor{\epsilon}{_b^\nu^\rho} \partial_\rho c^b_\nu \nonumber \\
&-\frac{1}{2} \partial_\mu (\phi + \pi) \partial^\mu (\phi + \pi) - \frac{1}{2} \Big( (\partial_\mu (V+W)_\nu)^2 - (\partial_\lambda (V+W)^\lambda)^2 \Big)\,.
\end{align}
Note that both, $d^a_\mu$ and $c^a_\mu$,  act as  Lagrange multipliers enforcing the respective constraints 
on each other,  as well as vanishing of the first two terms in the above Lagrangian. We are thus left with two free ordinary propagating fields in the decoupling limit, one scalar, $\pi + \phi$, and one vector, $V^a + W^a$.

\section{\label{sec:level1} Outlook}

We have found, not unexpectedly,  that the low energy dynamics of a fluid or solid coupled to gravity can be viewed as a general Lorentz violating massive gravity theory and that such a theory is free of ghosts with well behaved tensor, vector, and scalar modes. While these results could potentially have relevance to cosmology, the effects of such Lorentz violating metric perturbations  at shorter scales are better thought in terms of massless gravitons and matter fluctuations.
However, these results could be potentially interesting to study through the lens of AdS/CFT. The conformal field theory dual to massive gravity in various contexts has been studied in recent years \cite{Alberte2016, Alberte2017i, Alberte2017ii}. However, Lorentz violating massive gravity has not been studied in this context and the fact that an ordinary fluid/solid coupled to gravity gives rise to such a massive gravity theory could have interesting implications for the dynamics and symmetries of a corresponding dual CFT. Finally, in the Chern-Simons massive gauge theory studied in this paper, it's interesting that such an unusual mass term contracting spacetime with internal indices leads to a theory with well-behaved dynamics and stability, at least at low energies. It would be  interesting to see if there exist other such Chern-Simons gauge theories in diverse dimensions with gauge groups that allow for such special mass terms. In the (2+1)-d case, perhaps such effective field theories could show up in real condensed matter systems.

\section{\label{sec:level1} Acknowledgments}

GG would like to thank Oriol Pujolas for collaboration at an early stage of the project. The work
was supported in part by NSF grant PHY-1620039. 

\bibliographystyle{hunsrt.bst}
\bibliography{paper}

\begin{thebibliography}{10}

\bibitem{Anderson1963}
P.~W. Anderson.
\newblock Plasmons, gauge invariance, and mass.
\newblock {\em Phys. Rev.}, $\bf 130$:439--442, Apr 1963.

\bibitem{EnglertBrout1964}
F.~Englert and R.~Brout.
\newblock Broken symmetry and the mass of gauge vector mesons.
\newblock {\em Phys. Rev. Lett.}, $\bf 13$:321--323, Aug 1964.

\bibitem{Higgs1964}
Peter~W. Higgs.
\newblock Broken symmetries and the masses of gauge bosons.
\newblock {\em Phys. Rev. Lett.}, $\bf 13$:508--509, Oct 1964.

\bibitem{Kibble1964}
G.~S. Guralnik, C.~R. Hagen, and T.~W.~B. Kibble.
\newblock Global conservation laws and massless particles.
\newblock {\em Phys. Rev. Lett.}, $\bf 13$:585--587, Nov 1964.

\bibitem{Soper2008}
Davison~E. Soper.
\newblock {\em Classical Field Theory}.
\newblock Dover Publications, 2008.

\bibitem{Dubovsky2005}
S.~Dubovsky, T.~Gregoire, A.~Nicolis, and R.~Rattazzi.
\newblock Null energy condition and superluminal propagation.
\newblock {\em JHEP}, 03:025, 2006, hep-th/0512260.

\bibitem{Endlich2011}
Solomon Endlich, Alberto Nicolis, Riccardo Rattazzi, and Junpu Wang.
\newblock The quantum mechanics of perfect fluids.
\newblock {\em Journal of High Energy Physics}, 2011(4):102, Apr 2011,
  arXiv:hep-th/1011.6396.

\bibitem{Achucarro1986}
A.~Achucarro and P.~K. Townsend.
\newblock A chern-simons action for three-dimensional anti-de sitter
  supergravity theories.
\newblock {\em Phys. Lett.}, $\bf B180$:89, 1986.

\bibitem{WITTEN1986}
Edward Witten.
\newblock 2 + 1 dimensional gravity as an exactly soluble system.
\newblock {\em Nuclear Physics B}, $\bf 311$(1):46 -- 78, 1988.

\bibitem{Dunne1999}
G.~V. Dunne.
\newblock {\em Aspects Of Chern-Simons Theory}, pages 177--263.
\newblock Springer Berlin Heidelberg, Berlin, Heidelberg, 1999.

\bibitem{Rubakov2004}
V.~{Rubakov}.
\newblock {Lorentz-violating graviton masses: getting around ghosts, low strong
  coupling scale and VDVZ discontinuity}.
\newblock July 2004, arXiv:hep-th/0407104.

\bibitem{Dubovsky2004}
Sergei~L. Dubovsky.
\newblock Phases of massive gravity.
\newblock {\em Journal of High Energy Physics}, 2004($\bf 10$):076, 2004,
  arXiv:hep-th/0409124.

\bibitem{dRG}
C.~de~Rham and G.~Gabadadze.
\newblock Generalization of the fierz-pauli action.
\newblock {\em Phys. Rev. D}, $\bf 82$:044020, 2010, arXiv:hep-th/1007.0443.

\bibitem{dRGT}
G.~Gabadadze C.~de Rham and A.~J. Tolley.
\newblock Resummation of massive gravity.
\newblock {\em Phys. Rev. Lett.}, $\bf 106$:231101, 2011,
  arXiv:hep-th/1011.1232.

\bibitem{Esposito2017}
A.~Esposito, S.~Garcia-Saenz, A.~Nicolis, and R.~Penco.
\newblock Conformal solids and holography.
\newblock {\em Journal of High Energy Physics}, 2017(12):113, Dec 2017.

\bibitem{wald1984}
Robert~M. Wald.
\newblock {\em General relativity}.
\newblock The University of Chicago Press, 1984.

\bibitem{Alberte2016}
Lasma Alberte, Matteo Baggioli, Andrei Khmelnitsky, and Oriol Pujol{\`a}s.
\newblock Solid holography and massive gravity.
\newblock {\em Journal of High Energy Physics}, $\bf 2016$(2):114, Feb 2016,
  arXiv:hep-th/1510.09089.

\bibitem{Alberte2017i}
Lasma Alberte, Martin Ammon, Matteo Baggioli, Amadeo Jiménez-Alba, and Oriol
  Pujolàs.
\newblock {Holographic Phonons}.
\newblock 2017, 1711.03100.

\bibitem{Alberte2017ii}
Lasma Alberte, Martin Ammon, Matteo Baggioli, Amadeo Jiménez, and Oriol
  Pujolàs.
\newblock {Black hole elasticity and gapped transverse phonons in holography}.
\newblock {\em JHEP}, 01:129, 2018, 1708.08477.

\end{thebibliography}

\end{document}